\documentclass[%
 aip,
 amsmath,amssymb,
 reprint,%
]{revtex4-1}

\usepackage{graphicx}
\usepackage{dcolumn}
\usepackage{bm}

\usepackage[utf8]{inputenc}
\usepackage[T1]{fontenc}
\usepackage{mathptmx}
\usepackage{etoolbox}

\usepackage{amsfonts, amsmath, amsthm, amssymb}
\usepackage{graphicx}
\usepackage{listings}

\usepackage{amssymb, amsmath, amsbsy}
\usepackage{stackrel}

\usepackage{graphicx}
\usepackage{listings}
\usepackage{csquotes}
\usepackage[hidelinks]{hyperref}
\usepackage[nottoc,notlot,notlof]{tocbibind}

\usepackage{subcaption}
\usepackage[margin=1in]{geometry}
\usepackage{xcolor}
\usepackage{braket}

\makeatletter
\def\@email#1#2{%
 \endgroup
 \patchcmd{\titleblock@produce}
  {\frontmatter@RRAPformat}
  {\frontmatter@RRAPformat{\produce@RRAP{*#1\href{mailto:#2}{#2}}}\frontmatter@RRAPformat}
  {}{}
}%

\newcommand{\MH}[1]{#1}

\makeatother
\begin{document}

\preprint{AIP/123-QED}

\title[Oscillation Criteria in Large-Scale Gene Regulatory Networks with Intrinsic Fluctuations]{Oscillation Criteria in Large-Scale Gene Regulatory Networks with Intrinsic Fluctuations}

\author{Manuel Eduardo Hernández-García}
\homepage{manuel.hernandezgarcia@viep.com.mx}

\affiliation{ Facultad de Ciencias Físico Matemáticas, Benemérita Universidad Autónoma de Puebla, Heroica Puebla de Zaragoza 72570, México.
}%

\author{Jorge Velázquez-Castro}%
\homepage{jorge.velazquezcastro@correo.buap.mx}
\affiliation{ Facultad de Ciencias Físico Matemáticas, Benemérita Universidad Autónoma de Puebla, Heroica Puebla de Zaragoza 72570, México.
}%
 
\date{April 2026}

\begin{abstract}
Gene Regulatory Networks (GRNs) with feedback are essential components of many cellular processes and may exhibit oscillatory behavior. Analyzing such systems becomes increasingly complex as the number of components increases. Since gene regulation often involves a small number of molecules, fluctuations are inevitable. Therefore, it is important to understand how fluctuations affect the oscillatory dynamics of cellular processes, as this will allow comprehension of the mechanisms that enable cellular functions to remain even in the presence of fluctuations or, failing that, to determine the limit of fluctuations that permits various cellular functions.
In this study, we investigated the conditions under which GRNs with feedback and intrinsic fluctuations exhibit oscillatory behavior. Our focus was on developing a procedure that would be both manageable and practical, even for extensive regulatory networks, that is, those comprising numerous nodes.
Using the second-moment approach, we described the stochastic dynamics through a set of ordinary differential equations for the mean concentration and its second central moment. 
The system can attain either a stable equilibrium or oscillatory behavior, depending on its scale and, consequently, the intensity of fluctuations. To illustrate the procedure, we analyzed two relevant systems: a repressilator with three nodes and a system with five nodes, both incorporating intrinsic fluctuations. In both cases, it was observed that for very small systems, which therefore exhibit significant fluctuations, oscillatory behavior is inhibited. The procedure presented here for analyzing the stability of oscillations under fluctuations enables the determination of the critical minimum size of GRNs at which intrinsic fluctuations do not eliminate their cyclical behavior. 
\end{abstract}

\maketitle

\section{Introduction}\label{sec1}
Limit cycles and oscillations are fundamental features of numerous biological systems, including circadian rhythms \cite{Vitaterna, Mills}. In this study, we focused specifically on gene regulatory networks (GRNs) with negative feedback \cite{Schiavon}.  Deterministic models of these networks may have limit cycles and thus exhibit oscillatory behavior if delays or intermediate processes are present \cite{Xiao, Takada}. Within these systems, the cyclical dynamics of protein concentrations are pivotal for gene modules to perform their functions.

GRNs play crucial roles in cellular functions and development, enabling cells to respond to environmental stimuli by regulating gene expression \cite{Walk}. Examples of such systems include the p53–Mdm2 feedback loop \cite{Lev, Extri}, where p53 acts as a tumour suppressor and the mutations that are frequently found in cancer \cite{Smeenk, Baroni}. The synthesis of Hes-1, a transcription factor involved in stem cell maintenance, differentiation, and the inhibition of cancer progression \cite{Hirata, Liu}, is also an example of a gene module with a feedback loop.

However, these networks operate in regimes characterized by low molecule numbers and are thus significantly affected by stochastic fluctuations. In this study, we focus only on intrinsic fluctuations arising from their discrete nature and the randomness of molecular interactions \cite{Alon, VecchioM, Gar}. Several formalisms exist to study such stochastic systems, including the chemical master equation \cite{Gar}, stochastic simulation algorithms such as Gillespie’s algorithm \cite{Gillespie}, and approximate approaches such as the Langevin equation \cite{VecchioM, Langevin} and Fokker-Planck equations \cite{Scott}. An alternative is the moment approach \cite{Exact, Gomez, Manuel}, particularly the second-moment approach, which describes the dynamics of the system via a system of ordinary differential equations (ODEs) for the mean concentrations and their second central moments. Higher-order moments can be included as needed to capture nonlinear effects, such as those arising from Hill-type interactions \cite{Lakatos}. Furthermore, Hernandez et al. \cite{Exact} show that the dynamics of chemical networks consisting of first-order reactions or even nonlinear reaction rates such as Hill functions are described exactly by this method. In contrast with other methods, this ODE-based formulation enables the direct application of dynamical-systems theory, such as stability and bifurcation analysis \cite{Stability}, to stochastic systems, including results that classify system behavior in terms of equilibrium points or periodic orbits \cite{Smith}.

The goal of this work is to investigate large-scale GRNs with negative feedback under stochastic conditions driven by intrinsic fluctuations \cite{Kulasiri, Ribeiro} and to identify the conditions, including minimum protein concentrations, under which such systems can exhibit oscillatory behavior. Analyzing large chemical networks, as GRNs normally are, is particularly challenging due to the complexity of determining oscillation criteria in high-dimensional spaces. Prior studies employed control-theory graphical tools, based on the transfer function, to address these challenges in deterministic contexts \cite{Harat, Hori}. However, because the moment approach reformulates a stochastic system as a set of ODEs, in the present analysis, these methods are adapted to assess oscillatory conditions in stochastic systems.

We illustrate the methodology using two representative systems: a repressilator, a synthetic oscillator originally described by Elowitz et al. \cite{Elowitz} and previously studied primarily in deterministic frameworks \cite{Hanna}, and a five-node network consisting solely of repressors \cite{Meyer}. These examples show the applicability of the framework to cyclic GRNs with negative feedback and an arbitrary number of nodes and highlight the impact of system size and component concentrations on the emergence and attenuation of oscillations.  

The remainder of this paper is organized as follows. In Section \ref{section2}, we present the chemical master equation and a set of ODEs for the mean and second central moments using the moment approach. In Section \ref{section3}, we analyze a cyclic GRN with negative feedback, and we identify the conditions under which oscillations appear. In Section \ref{section4}, we apply the framework to two representative systems and examine the impact of intrinsic fluctuations on the 
emergence of oscillations. Finally, in Section \ref{section5} we present our results and conclusions.

\section{Moment Approach} \label{section2}  

In this work, we consider only intrinsic fluctuations; for this purpose, we followed the methodology in \cite{Gar} to derive the Chemical Master Equation (CME).  Let be $N$ chemical species, $S_l$ ($l$ $\in$ \{$1,2,..N$\}) and $r$ reactions $\mathcal{R}_i$ ($i$ $\in$ \{$1,2,...,r$\}) where species are transformed as follows: 
\begin{align}   
\mathcal{R}_i : \sum_{l=1}^{N} \alpha_{il} S_l \stackbin[]{k_{i}}{\rightarrow} \sum_{l=1}^{N} \beta_{il} S_l. 
\end{align}
$k_i$ is the kinetic parameter of the reaction rate, and coefficients $\alpha_{il} $ and $\beta_{il}$ are non-negative integers. From these, we derive the stoichiometric matrix of the system $ \Gamma_{il}= \beta_{li} -  \alpha_{li} $ (indices are exchanged because it is the transpose). Through collisions (or interactions) between different elements, the system evolves according to the law of mass action and the propensity rates given by  \cite{Gar}:  
\begin{align}    
{a_i(\mathbf{S})}&= k_{i} \prod _{l=1}^N \frac{S_l !}{\Omega ^{\alpha_{il}}(S_l- \alpha_{il} )!},   
\end{align}   
where the index $i$ corresponds to reactions $\mathcal{R}_i$ and $\mathbf{S}= (S_1, S_2, ..., S_{N})$, the vector of number of chemical species. These propensities represent the transition probabilities per unit time between different system states, where $\Omega = N_{A}V$ is the size of the system and has units of $volume/mol$, and Avogadro's number $N_{A}$ is used to convert the number of molecules to moles and has units of $1/mol$ \cite{Decimal}. We now obtain the CME: 
{\small
\begin{align}  
\partial_{t} P(\mathbf{S},t)= \Omega  \sum_{i=1}^{r} \left( a_{i}(\mathbf{S}-\Gamma_{i}) P(\mathbf{S}-\Gamma_{i},t) - a_{i}(\mathbf{S}) P(\mathbf{S},t) \right), \label{3}  
\end{align} } 
where $\Gamma_{i}$ represents the $i$-th column of matrix $\bm\Gamma$. The CME describes the temporal evolution of the system's probability states.  Analyzing the conditions of oscillations directly from the CME (\ref{3}) is extremely difficult. Therefore, we use the second-moment approach \cite{Gomez, Manuel}, which transforms the CME into a set of ordinary differential equations (ODEs). This formulation allows for stability analysis, which is well established for ODE systems, and can be applied to chemical processes that exhibit intrinsic fluctuations \cite{Stability}. 

To obtain the second-moment approach, we first define the mean concentration of chemical species $l$, $s_l = \frac{\langle S_l \rangle}{\Omega}$. The second central moment between chemical species $l_1$ and $l_2$ ($l_1,l_2$ $\in$ \{$1,2,..N$\});  $M^2_{l_1,l_2}= \frac{1}{\Omega^2} \braket{(S_{l_1}- \braket{S_{l_1}})(S_{l_2}- \braket{S_{l_2}})}$.  We assume that any analytical function  $F(\mathbf{S})$  of the system variables can be expanded using a Taylor expansion around the mean concentration, as shown below:   
\begin{widetext}
{\small 
\begin{align}   
\braket{F(\mathbf{S})} \approx & \left \langle F( \braket{\mathbf{S}}) +  \sum_{l_1} (S_{l_1}- \braket{S_{l_1}})\frac{\partial F( \braket{\mathbf{S}})}{\partial S_{l_1}} + \sum_{l_1,l_2}  \frac{(S_{l_1}- \braket{S_{l_1}})(S_{l_2}- \braket{S_{l_2}})}{2} \frac{\partial^2 F( \braket{\mathbf{S}})}{ \partial S_{l_1} \partial S_{l_2}} \right \rangle =F( \braket{\mathbf{S}}) + \sum_{l_1,l_2}  \frac{M^2_{l_1,l_2} \Omega^2}{2} \frac{\partial^2 F( \braket{\mathbf{S}})}{ \partial S_{l_1} \partial S_{l_2}}, \label{5} 
\end{align}}
\end{widetext}
where $\braket{\mathbf{S}}= (\braket{S_1},\braket{S_2},...,\braket{ S_N})$  is the mean state vector.  If the function $F(\mathbf{S})$ is a polynomial of order 2, then the expansion in Equation (\ref{5}) is exact.  

To derive the ODEs for the mean concentration and the second central moments from the CME, we considered only reactions up to first order. For the mean concentration, we multiplied the CME by $S_l$, computed the mean, and utilized (\ref{5}). A similar approach was applied to the second central moments, resulting in the following equations:

\begin{subequations} \label{6}
{\small
\begin{align}    
\frac{\partial s_l }{\partial{t}} & =  \sum_{i=1}^r \Gamma_{li} k_iR_i(\mathbf{s}) ,  \\ \frac{\partial M^2_{{l_1}, {l_2}}}{\partial{t}}  & = \sum_{i=1}^r \left( \frac{\Gamma_{l_1 i} \Gamma_{l_2 i} k_iR_i(\mathbf{s})}{\Omega}     \right.   \\
& \left. +\sum_{j_1=1}^{N} \left( M^2_{{l_1}, {j_1}} \Gamma_{l_2 i}k_i \frac{\partial R_i(\mathbf{s})}{\partial {s_{j_1}}}  + M^2_{{j_1}, {l_2}} \Gamma_{l_1 i} k_i \frac{\partial R_i(\mathbf{s})}{\partial {s_{j_1}}} \right) \right) \nonumber,  
\end{align}} 
\end{subequations}
where $\mathbf{s}=(s_1,s_2,..., s_N)$  and $R_i(\mathbf{s}) =  \prod_{j=1}^{N} s_j^{\alpha_{ij}}$, $k_iR_i(\mathbf{s})$ are the reaction rates; then, we can describe the system exactly via a set of ODEs.

Following Corollary 1 in \cite{Exact}, where there are only zero- and first-order reactions and functional kinetic parameters $k_i=k_i^* f_i(\mathbf{s}, \mathbf{M}^2)$, with $f_i$ depending on the mean concentration and the second central moment (for example, Hill functions), we can describe the system exactly considering up to the second central moment as in Equation \eqref{6}. Now we can explore the stability \cite{Stability} and establish the oscillation conditions in a system with intrinsic fluctuations.

\section{Gene Regulatory Network with Negative Feedback} \label{section3} 

\begin{figure}[h!t]
    \centering
    \includegraphics[width=0.85\linewidth]{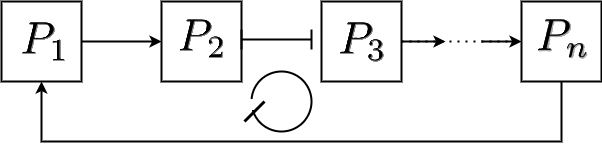}
    \caption{\textbf{Cyclic gene regulatory network.} In this figure, there is a cyclic gene regulatory network with negative feedback, where the arrows represent positive regulation, and the bars represent negative regulation, but the complete cycle has negative feedback. At each node, there is a module of transcription-translation processes that synthesizes a protein that acts as a transcription factor for the next module. }
    \label{fig.1}
\end{figure}

Gene regulatory networks (GRNs) are fundamental to biology and play a crucial role in cellular processes and development.  These networks enable cells to respond to environmental changes by coordinating their gene expression \cite{Walk}. In particular, GRNs with negative feedback,  in the deterministic description, where there are no fluctuations, can exhibit oscillations if these systems have a delay, or intermediate processes \cite{Xiao, Takada}. We then explore whether these types of systems exhibit oscillations. Some important systems in which these appear are the p53–Mdm2 feedback loop \cite{Lev, Extri}, where p53 is a tumor suppressor \cite{Smeenk, Baroni}, and the synthesis of Hes-1, in which Hes-1 plays a crucial role in various cellular processes, including stem cell maintenance, differentiation, and cancer progression \cite{Hirata, Liu}. 

In this study, we utilized the cycle gene regulatory network depicted in Figure \ref{fig.1}, where each node features a transcription–translation module that produces a protein functioning as a transcription factor with repressive or activating roles described by Hill functions, contributing to an overall negative-feedback mechanism. This system, proposed in prior studies \cite{Banks}, is extended here to investigate the impact of intrinsic fluctuations on oscillation emergence, building upon previous work by \cite{Harat, Hori}. We also used the Hill function with stochastic corrections; for more details, please refer to \cite{Exact}.

\subsection{Model}
 
We further analyzed the cyclic gene regulatory network in Figure \ref{fig.1}, where activating and repressing regulatory interactions are represented by arrows and bars, respectively, while emphasizing the network's negative-feedback structure across $n$ modules. Because each transcription-translation module has similar reactions, the reactions in each module are

\begin{table}[h!]
    \centering
    \begin{tabular}{|c|c|}
        $ \emptyset \stackbin{\gamma_{1,j}}{\longrightarrow} M_j$ & $M_j  \stackbin{\delta_{1}}{\longrightarrow} \emptyset$ \\
        $M_j    \stackbin{\gamma_{2,j}}{\longrightarrow} M_j + P_j  $ & $ P_{j}  \stackbin{\delta_{2}}{\longrightarrow} \emptyset $ \\
    \end{tabular}
\end{table}
\noindent where $j \in (1,2,...,n)$. The first reaction is the synthesis of the mRNA by a functional \MH{kinetic} parameter $\gamma_{1,j}$ (in this case a Hill function) and its degradation \MH{depends on the kinetic parameter} $\delta_1$. The next reaction is the synthesis of the proteins by the parameter $\gamma_{2,j}$   and a degradation \MH{depends on the kinetic parameter} $\delta_2$.  Note that the degradation parameters for the mRNA and the proteins are equal across all modules, respectively.  

The stoichiometric coefficients in module $j$ are $\alpha_{il}^j$ and $\beta_{il}^j$, and the stoichiometric matrix are,

{\small
\begin{align}
 \alpha_{il}^j&= \begin{pmatrix}
		0& 0   \\
            1& 0   \\
            1& 0   \\
            0& 1   \\
	\end{pmatrix} ,    &
 \beta_{il}^j&=  \begin{pmatrix}
		1& 0   \\
            0& 0   \\
            1& 1   \\
            0& 0   \\
	\end{pmatrix},  &
 \Gamma_{li}^j&= \begin{pmatrix}
		1& -1 & 0 & 0   \\
            0 & 0 & 1& -1      \\
	\end{pmatrix}. \nonumber
\end{align}}

The reaction rates in module $j$ are,
\begin{align}
   R_{1,j}&= 1  , &
    R_{2,j}&=  m_j, \nonumber \\
   R_{3,j}&= m_j, &
    R_{4,j}&=  p_j,
\end{align}
where $m_j$ and $p_j$ are the mean concentrations of the mRNA and protein in module $j$. If the system has $n$ nodes, then the system has $2n + 4$ kinetic parameters, but if we dimensionally reduce the system (see Appendix \ref{A.A} for details), the number is reduced to $n+2$. Then without loss of generality, we choose $\gamma_{2,j}=\delta_2=1$ and $\delta_1=\delta$ and $\gamma_{1,j}= \delta \gamma_j^2 H_j$, and $H_j$ is defined by, 
\begin{widetext}
\begin{align}
    H_j= H_j(p_{j-1},M^2_{p_{j-1},p_{j-1} } )=  \begin{cases}         
 \left(  \frac{1}{1+ p^2_{j-1} + M^2_{p_{j-1},p_{j-1} } - \frac{p_{j-1}}{\Omega} }\right) & \text{for a repressor } \\       
 \left(  \frac{p^2_{j-1} + M^2_{p_{j-1},p_{j-1} }  - \frac{p_{j-1}}{\Omega}}{1+ p^2_{j-1} + M^2_{p_{j-1},p_{j-1} } - \frac{p_{j-1}}{\Omega} }\right) & \text{for an activator} 
\end{cases}  , 
\end{align}
\end{widetext}
where $M^2_{p_j,p_j} $ is the second central moment of the protein $P_j$. These are the Hill functions for a repressor or activator, respectively, in which we used a Hill coefficient equal to two and a dissociation constant equal to one. \MH{For more details on the derivation, see Appendix \ref{A.0}. The Hill function captures the fast binding of the transcription factors to the promoter region. Specifically, the derivation relies on a separation of timescales, in which the binding and unbinding of transcription factors occur much faster than the dynamics of mRNA and protein. This justifies a quasi-equilibrium approximation for the promoter states. Deviations from this assumption may compromise the validity of the Hill-function approximation. In such cases, it is more appropriate to explicitly model the underlying biochemical reactions from which the Hill function is derived. } This system can be described exactly up to the second central moment because there are only first-order reactions and functional kinetic parameters, that is, the Hill functions \cite{Exact}. The ODES for the mean concentrations and the second central moment are:
\begin{subequations} \label{8}
\begin{align}
    \frac{\partial m_j}{ \partial t}=& \delta( \gamma_j^2 H_j - m_j),  \\
    \frac{\partial p_j}{ \partial t}=&  m_j - p_j , \\
    \frac{\partial M^2_{m_j,m_j}}{ \partial t}=& \frac{\delta}{\Omega} \left( \gamma_j^2 H_j +  m_j \right) - 2 \delta M^2_{m_j,m_j}, \\
    \frac{\partial M^2_{m_j,p_j}}{ \partial t}=& M^2_{m_j,m_j} - (\delta + 1) M^2_{m_j,p_j}, \\
    \frac{\partial M^2_{p_j,p_j}}{ \partial t}= &\frac{1}{\Omega} \left( m_j +  p_j \right)+ 2 M^2_{m_j,p_j} - 2 M^2_{p_j,p_j} . 
\end{align}
\end{subequations}

Equations \eqref{8} are a set of ODEs and provide an accurate description of the system dynamics. In the next subsection, we will perform the stability analysis directly on these equations.

\subsection{Existence of Oscillations}

In the deterministic description, where fluctuations are not considered, the GRNs with negative feedback may exhibit oscillations \cite{Xiao, Takada}. However, when intrinsic fluctuations are considered, it is not straightforward to determine their appearance, especially for large GRNs. On the other hand, in a previous work \cite{Stability}, it was shown that stability analysis can be performed on a stochastic system with intrinsic fluctuations using the moment approach, yielding a set of ODEs.

In the study of differential equations, the Poincaré–Bendixson theorem is a fundamental tool for proving the existence of limit cycles (sustained oscillations) in nonlinear dynamical systems. Although originally formulated for systems in $\mathbb{R}^2$, generalizations exist for certain higher-dimensional systems with special structures, such as monotonic cyclic networks \cite{Mallet}. In our case, the system is feedback-dependent on two variables: the mean concentration and the second central moment of the proteins. Despite the complexity, the system is still represented by a sets of ODEs, and we can base the conjecture given by \cite{Smith, Mallet} for higher dimensions and based on the results of \cite{Hori} we suppose that the set of ODEs in \eqref{8} has only two types of dynamics: it tends to an equilibrium point or a periodic orbit (oscillations).

We can analyze the stationary state, then we have the following values for the second central moment,
\begin{subequations}\label{10}
\begin{align}
    M^2_{m_j,m_j,ss}=& \frac{m_{j,ss}}{\Omega},  \\
     M^2_{m_j,p_j,ss}=& \frac{m_{j,ss}}{\Omega} \left( \frac{1}{\delta + 1} \right) ,  \\
     M^2_{p_j,p_j,ss}=& \frac{p_{j,ss}}{\Omega}\left( 1+ \frac{1}{\delta + 1}\right).  
\end{align}
\end{subequations}
From these results, we can say that the mRNA has a Poisson distribution in the stationary state, whereas the proteins have \MH{an over-dispersion in comparison with a Poisson distribution}. For the mean concentrations of proteins, we have 
\begin{align}
    p_{j,ss}=  m_{j,ss}. \label{11}
\end{align}
To obtain the values of the mean concentrations of the mRNA, we need to solve the following equations

\begin{align}
    m_{j,ss} = \gamma_j^2 \times \begin{cases}   
     \left(  \frac{1}{1+ m^2_{j-1,ss} + \frac{m_{j-1,ss}}{\Omega}\left( \frac{1}{\delta + 1}\right) }\right) & \text{for a repressor } \\  
\left(  \frac{m^2_{j-1,ss} + \frac{m_{j-1,ss}}{\Omega}\left( \frac{1}{\delta + 1}\right)}{1+ m^2_{j-1,ss} + \frac{m_{j-1,ss}}{\Omega}\left( \frac{1}{\delta + 1}\right) }\right) & \text{for an activator} 
\end{cases}. \label{12}
\end{align}

To conduct the analysis in the upcoming sections, we'll need the derivatives of the Hill function evaluated at the stationary point shown below.
\begin{widetext}
\begin{subequations}
\begin{align}
    \Xi_j= & \left. \frac{\partial  H_j}{ \partial p_{j-1}} \right|_{ss} = \left(  \frac{2 m_{j-1,ss} - \frac{1}{\Omega}}{ \left( 1+ m^2_{j-1,ss} + \frac{m_{j-1,ss}}{\Omega}\left( \frac{1}{\delta + 1}\right) \right)^2 }\right) \times\begin{cases}         
  -1 & \text{for a repressor } \\  
  +1 & \text{for an activator}    
\end{cases},  \\
    \xi_j=  & \left. \frac{\partial  H_j}{ \partial M^2_{p_{j-1},p_{j-1}}} \right|_{ss} = \left(  \frac{1}{ \left( 1+ m^2_{j-1,ss} + \frac{m_{j-1,ss}}{\Omega}\left( \frac{1}{\delta + 1}\right) \right)^2 }\right) \times \begin{cases}         
  -1 & \text{for a repressor } \\  
  +1 & \text{for an activator}   
\end{cases}.
\end{align}
\end{subequations}
\end{widetext}
In the next subsection, we present a method to determine the appearance of stable oscillations. \MH{To has positive values between the parenthesis $2m_{j,ss}>1/\Omega$.}

\subsection{Transfer Function and Criteria of Oscillation}

Then, to determine the occurrence of stable oscillations, we need to find the eigenvalues of the system at the stationary state. If there is an eigenvalue with a negative real part, the system is stable and converges to the equilibrium point; otherwise, oscillations (a periodic orbit) appear. However, for this type of system, there are $5n$ eigenvalues, which increase as the number of nodes $n$ increases. Thus, for a system with a large $n$, it is difficult to determine all eigenvalues. To address this issue and determine whether oscillations occur, we followed a procedure based on \cite{Harat, Hori}, where transfer functions and a stability criterion were used to develop a graphical criterion. 

To begin, we linearize the set of ODEs \eqref{8} around the equilibrium point to determine the system's transfer function. As the modules are alike, we can just analyze any module $j$. The following variables are introduced to carry out the linearization; $\Delta m_j= m_j - m_{j, ss}$, $\Delta p_j= p_j - p_{j, ss}$, $\Delta M^2_{m_j,m_j} = \Omega(M^2_{m_j,m_j}- M^2_{m_j,m_j, ss})$, $\Delta M^2_{m_j,p_j} = \Omega(M^2_{m_j,p_j}- M^2_{m_j,p_j, ss})$ and $\Delta M^2_{p_j,p_j} = \Omega(M^2_{p_j,p_j}- M^2_{p_j,p_j, ss})$. Note that we multiply the second central moment by $\Omega$; this is done to have a set of variables with values in the same regime and the same units. Then we treat the influence of the previous transcription–translation module as an applied control, $u_j$, and write the equations for each module $j$ of transcription–translation as, 

\begin{widetext}
\begin{align}
\frac{\partial}{\partial t}
\begin{pmatrix}
\Delta m_j \\
\Delta p_j \\
\Delta M^2_{m_j,m_j} \\
\Delta M^2_{m_j,p_j} \\
\Delta M^2_{p_j,p_j}
\end{pmatrix}
=&
\begin{pmatrix}
-\delta & 0 & 0 & 0 & 0 \\
1 & -1 & 0 & 0 & 0 \\
{\delta} & 0 & -2\delta & 0 & 0 \\
0 & 0 & 1 & -(\delta + 1) & 0 \\
1 & 1 & 0 & 2 & -2
\end{pmatrix}
\begin{pmatrix}
\Delta m_j \\
\Delta p_j \\
\Delta M^2_{m_j,m_j} \\
\Delta M^2_{m_j,p_j} \\
\Delta M^2_{p_j,p_j}
\end{pmatrix}
+ 
\begin{pmatrix}
\delta \\
0 \\
\delta \\
 0 \\
 0
\end{pmatrix} u_j  . \label{18}
\end{align}
\end{widetext}
These equations are written in the form $\frac{\partial}{\partial t} \mathbf{x}_j= \mathbf{A} \mathbf{x}_j + \mathbf{B} u_j$ with $u_j=\gamma_j^2( \Xi_j \Delta p_{j-1} + \frac{1}{\Omega} \xi_j \Delta M^2_{p_{j-1},p_{j-1}} )$. Then from Equation \eqref{18} we identify $\mathbf{A}$ and $\mathbf{B}$. Additionally, $\mathbf{A}$ and $\mathbf{B}$ are similar for all modules.

Now we write $u_j$ in a matrix form $\mathbf{u}= \mathbf{K}_1 \Delta\mathbf{p}+ \mathbf{K}_2 \Delta \mathbf{M}_{p,p}^2$
\begin{widetext}
{\footnotesize
\begin{align}
\begin{pmatrix}
u_1 \\
u_2 \\
u_3 \\
\vdots\\
u_n
\end{pmatrix}
= 
\begin{pmatrix}
0 & 0 & \dots & 0& \gamma_1^2 \Xi_1  \\
\gamma_2^2\Xi_2  & 0 &  \dots& 0 &0\\
0 & \gamma_3^2 \Xi_3  &  \dots &0 &0  \\
\vdots &  \vdots & \ddots &\vdots & \vdots \\
 0 &0 & \dots& \gamma_n^2 \Xi_n  & 0 \\
\end{pmatrix}
\begin{pmatrix}
\Delta p_1 \\
\Delta p_2 \\
\vdots \\
\Delta p_{n-1} \\
\Delta p_n \\
\end{pmatrix} 
+ 
\begin{pmatrix}
0 & 0 & \dots & 0& \frac{\gamma_1^2}{\Omega} \xi_1  \\
\frac{\gamma_2^2}{\Omega} \xi_2  & 0 &  \dots& 0 &0\\
0 &  \frac{\gamma_3^2}{\Omega}  \xi_3  &  \dots &0 &0  \\
\vdots &  \vdots & \ddots &\vdots & \vdots \\
 0 &0 & \dots& \frac{\gamma_n^2}{\Omega}  \xi_n  & 0 \\
\end{pmatrix}
\begin{pmatrix}
\Delta M^2_{p_1,p_1} \\
\Delta M^2_{p_2,p_2} \\
\vdots \\
\Delta M^2_{p_{n-1},p_{n-1}} \\
\Delta M^2_{p_n,p_n} \\
\end{pmatrix}, \label{14}
\end{align}} 
\end{widetext}
and define the matrix:
\begin{align}
    \mathbf{K}= \begin{pmatrix}
               \mathbf{K}_1  &
                \mathbf{K}_2
 \end{pmatrix}.
\end{align}

\MH{Each matrix $\mathbf{K}_1$ and $\mathbf{K}_2$ has $n \times n$ entries, and matrix $\mathbf{K}$ has $n \times 2n$ entries.  Note that $\mathbf{K}_1$ and $\mathbf{K}_2$ are very similar because both have a cyclic structure; however, in Equation $\mathbf{K}_2$ there is a factor $1/\Omega$, which decreases while $\Omega$ increases.} Now, we calculate eigenvalues of the matrix $\mathbf{K}_1$  then we obtained
\begin{align}
    \lambda_{j}=& L e^{I(2j-1) \pi/n}, L=  \left| \prod_{j=1}^{n} \gamma_j^2 \Xi_j \right|^{\frac{1}{n}}, \label{16} 
\end{align}
where $j \in (1,2,...,n)$ and $I$ is the imaginary unit. \MH{Note that a term $-1$ appears in the exponential, since the product of all $\prod_j \Xi_j$ is negative because the GRN has negative feedback.} 

The protein and its second central moment of a module $j$ affect the next module in the cycle, and then its values are the output in each module; for this, we have
\begin{align}
 \mathbf{y}_j=   \mathbf{Cx}_j= 
    \begin{pmatrix}
0 & 1 & 0 & 0 & 0 \\
0 &  0 & 0 & 0 & 1\\
\end{pmatrix}
\begin{pmatrix}
\Delta m_j \\
\Delta p_j \\
\Delta M^2_{m_j,m_j} \\
\Delta M^2_{m_j,p_j} \\
\Delta M^2_{p_j,p_j}
\end{pmatrix},
\end{align}

matrix $\mathbf{C}$ is the same for all modules. From $\mathbf{A}$, $\mathbf{B}$, and $\mathbf{C}$, we found that each module is observable but not controllable. Observability is attributed to the ability to infer the values of other variables from the output variables. The lack of controllability is due to intrinsic fluctuations within the system, which impede full control over its dynamics. In particular, it is not possible to control the variable $M^2_{m_j,p_j}$, which means that the interactions between the fluctuations of $m_j$ and $p_j$ cannot be controlled. Nevertheless, although the system is uncontrollable, the uncontrollable variable remains stable (for further details see Appendix \ref{B}). As a result, the overall stability of the system depends on the controllable variables.

Following the procedure to determine the stability conditions of the controllable variables\cite{Harat, Hori}, we now calculate the transfer function of each module,
\begin{align}
    \mathbf{g}(s)= \mathbf{C (}s\mathbf{I - A)^{-1} B}= 
    \begin{pmatrix}
g_1(s) \\
g_2(s)
\end{pmatrix},
\end{align}
where
\begin{subequations}\label{20}
{\small
\begin{align}
    g_1(s)=& \frac{1}{(T_as+1)(T_bs+1)},  \\
    g_2(s)=& \frac{1}{(T_as+1)(T_bs+1)} \left(  1 +      \frac{T_d(T_as+1)}{(T_cs+1)(T_ds+1)}  \right), 
\end{align}}
\end{subequations}

\begin{figure}[h!t]
    \centering
    \includegraphics[width=0.47\linewidth]{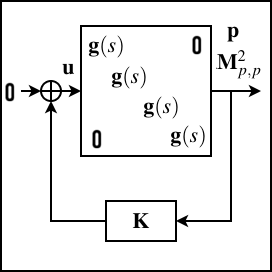}
    \includegraphics[width=0.47\linewidth]{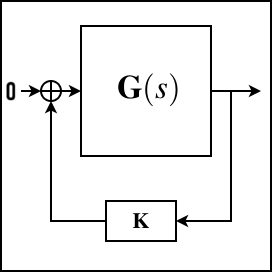}
    \caption{\textbf{Block diagram.} \MH{We show the overall GRN, composed of $n$ identical modules, each with internal dynamics given by $\mathbf{g}(s)$, an input $u$, and outputs $p$ and $M^2_{p,p}$. The interconnections between modules are captured as output feedback modulated by $K$. $\mathbf{G}(s)$ has $n$ inputs and $2n$ outputs.}      }
    \label{fig.1a}
\end{figure}

$T_a=1$, $T_b= \frac{1}{\delta}$, $T_c= \frac{1}{2}$ and $T_d= \frac{1}{1+\delta}$. The transfer function is the same for each module because $\mathbf{A}$, $\mathbf{B}$, and $\mathbf{C}$ are the same for each module. $g_1(s)$ is identical to that obtained in the deterministic description, but another transfer function, $g_2(s)$, appears here that is related to the second central moment. 

\MH{We define a transfer function of the overall GNR that separates the contribution of the first and the second output, that is
\begin{align}
    \mathbf{G}(s)= 
\begin{pmatrix}
g_1(s) I_{n\times n}\\
g_2(s) I_{n\times n}     
\end{pmatrix}, 
\end{align}
$\mathbf{G}(s)$ has $2n \times n$ entries. The original feedback system (left panel of Figure \ref{fig.1a}) can be transformed into an equivalent feedback representation \cite{Harat, Hori}, as shown in the right panel of Figure \ref{fig.1a}. Then, the overall transfer function of the system is  
\begin{align}
    \mathcal{H}(s)=  (\mathbf{I} - \mathbf{G}(s)\mathbf{K})^{-1} \mathbf{G}(s). \label{22}
\end{align}}
This expression corresponds to a class of large-scale linear systems with generalized frequency variables.  \MH{ $\mathcal{H}(s)$ can thus be viewed as an interconnection of $n$ identical modules, each of which has the internal dynamics $\mathbf{g}(s)$, where the interconnection structure is specified by $\mathbf{K}$, and the input-output structure for the whole system is specified by $\mathbf{B}$ and $\mathbf{C}$ \cite{Hara2013}.} To determine the stability of the system \cite{Harat}, we must calculate the poles of Equation \eqref{22}. For this, we calculate the determinant of the inverse part of this expression equal to zero, \MH{and to simplify, we use the Weinstein–Aronszajn identity (or Sylvester's determinant theorem) \cite{Pozrikidis} (for more details see Appendix \ref{D}), then,
\begin{align}
    0=& det (I_{2n \times 2n} - \mathbf{G}(s)\mathbf{K} ) \nonumber \\
    =& det (I_{n \times n} - \mathbf{K} \mathbf{G}(s)) \nonumber \\
    =&det( I_{n \times n} - g_1(s) \mathbf{K}_1 - g_2 \mathbf{K}_2), 
\end{align}
because $\mathbf{K}_1$ and $\mathbf{K}_2$ have a cyclic structure as can see in \eqref{14}, calculating the determinant, we have
{\small
\begin{align}
    0 =& 1- \prod_{j=1}^{n} \gamma_j^2 \left(  g_1(s)  \Xi_j + g_2(s) \frac{1}{\Omega}  \xi_j \right) \nonumber \\
    =&1- \left( \prod_{j=1}^{n} \gamma_j^2  \Xi_j \right) \left(  g_1^n(s) \prod_{j=1}^{n} \left(1  +  \frac{1}{\Omega} \frac{g_2(s)}{g_1(s)}  \frac{\xi_j}{\Xi_j} \right)\right).
\end{align}} 
After a bit of algebra, we get 
\begin{align}
    0=&\phi^n(s)- \prod_{j=1}^{n} \gamma_j^2 \Xi_j, \label{25.a} 
\end{align}
where,
\begin{align}
    \phi(s)= & \frac{1}{g_1(s)} \left( \prod_{j=1}^n \left( 1 + \frac{1}{\Omega} \frac{g_2(s)} {g_1(s)} \frac{\xi_j}{\Xi_j}\right) \right)^{-\frac{1}{n}}. \label{25.b}
\end{align}}
In the case in which $\Omega \to \infty$ (\MH{ the deterministic limit}) the Equation \eqref{25.b} is reduced to $\phi(s)=1/g_1(s)$. This is the expression obtained in the deterministic description of the system \cite{Hori}. \MH{We can consider the term between the parentheses that depends of $1/\Omega$ as the stochastic correction. } In the following, we consider the full Equation \eqref{25.b} to assess the system's stability. The system is unstable if there exists any value of $s$ with  $Re(s)\geq 0$ such that \eqref{25.a} is satisfied\cite{Harat}. Instability implies that there are oscillations in the system. We formalize this result in the following proposition:\\

\textbf{Proposition 1. \label{P.2}} 
Consider a cyclic gene regulatory network with dynamics \eqref{8} and its linearized system in (\ref{22}). Then the system has oscillations if there exists any value of $s$ with $Re(s)> 0$ such that

    \begin{equation}
         \phi^n(s) - \prod_{j=1}^{n} \gamma_j^2 \Xi_j  = 0. \label{31}
    \end{equation}

We can then apply and extend the results of \cite{Harat, Hori} to obtain a condition in which we can visualize whether a cycle gene regulatory network has oscillations. Before continuing, we define the next space $\mathbb{C}_+= \{  s \in \mathbb{C}_{+}| Re(s) > 0  \}$, that is, the set of a complex number with a positive real part.  \\

\textbf{Proposition 2. \label{P.3}}    
Consider a cyclic gene regulatory network with dynamics \eqref{8} and its linearized system in (\ref{22}). 

\begin{itemize}
    \item Then the system has oscillations if at least one of the eigenvalues of $\mathbf{K}_1$, $\lambda$, exists within the domain $\Lambda^+$, defined by $\Lambda^+= \phi(\mathbb{C}_+)$. 

    \item Otherwise, the system converges to a steady state. 
\end{itemize}

Note that the set of values of $\Lambda^+$ is defined by evaluating $\phi(s)$ using complex values from $\mathbb{C}_+$, when at least one eigenvalue of matrix $\mathbf{K}_1$ is in this region, the system oscillates. This proposition addresses the effects of fluctuations and system size on the emergence of oscillations. When the size of the system increases, the result in \cite{Hori} is recovered. \MH{ A sketch of the proof follows from Proposition 1, where the $n$-th root is taken on Equation \eqref{31}. }

This graphical criterion implies that the stability of $\mathcal{H}(s)$ can be easily assessed by examining the eigenvalue distribution of the matrix $ \mathbf{K}_1$ describing protein interactions and the domains $\Lambda^+$ defined by the homogeneous (ARm/protein) dynamics $\phi(s)$\cite{Hori}. This criterion reduces the complexity of determining system stability, even when $n$ is large. Next, we use Proposition 2 to analyze some systems.

\section{Examples} \label{section4}

\subsection{Repressilator}

\begin{figure}[h!t]
    \centering
    \includegraphics[width=0.4\linewidth]{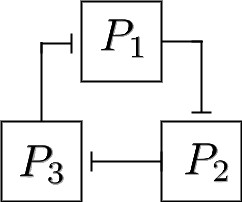}
    \caption{\textbf{Repressilator.}  This is a cycle gene regulatory network with three modules, in which the protein represses mRNA synthesis in the next module, and the system has negative feedback.  }
    \label{fig.2}
\end{figure}

The first example we analyze is the repressilator, as shown in Figure \ref{fig.2}, which consists of three modules: each protein represses the synthesis of the next, and its dynamics are described by Equations \eqref{8}. This gene regulatory network was developed experimentally by \cite{Elowitz}. In previous work, this system was analyzed \cite{Hori, Hanna}, and the criteria for oscillations were determined without considering fluctuations. However, here, we extend this to include intrinsic fluctuations based on the previous section.

\begin{table}[htbp]
\centering
\begin{tabular}{cccc}
\hline
\textbf{Variable} & \textbf{Values} & $\Omega$ \\
\hline
$\mathbf{m}_{ss}$ & $(1.145, 1.388, 2.210)$  & 1 \\
$\mathbf{m}_{ss}$ & $(1.237, 1.543, 2.313)$  & 10 \\
$\mathbf{m}_{ss}$ & $(1.248, 1.560, 2.324)$  & 100 \\
\hline
\end{tabular}
\caption{\textbf{Stationary mean concentrations of the repressilator.} The stationary value of the mean concentration of the mRNA obtained with Equation \eqref{12} using different values of $\Omega$.  }
\label{tabla1}
\end{table}

\begin{figure*} [h!t]
  \begin{subfigure}{\linewidth}
\includegraphics[width=.33\textwidth]{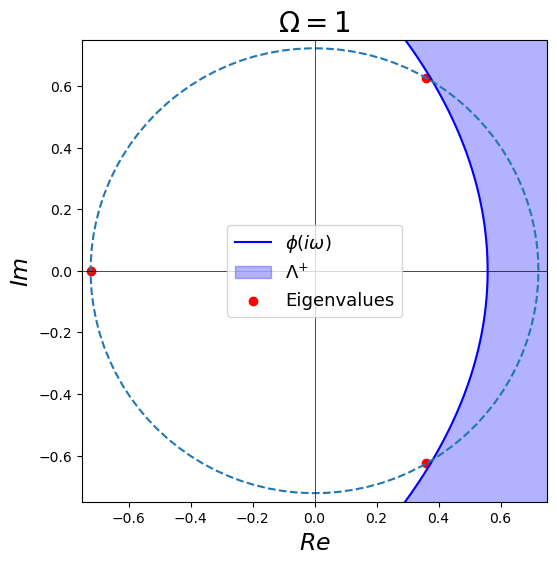}\hfill
\includegraphics[width=.33\textwidth]{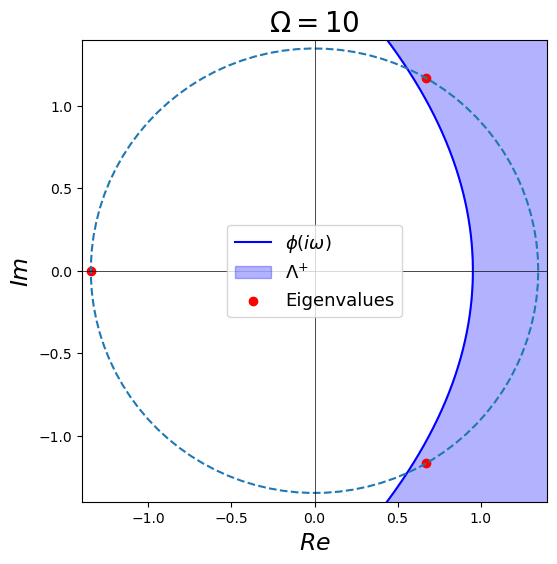}\hfill
\includegraphics[width=.33\textwidth]{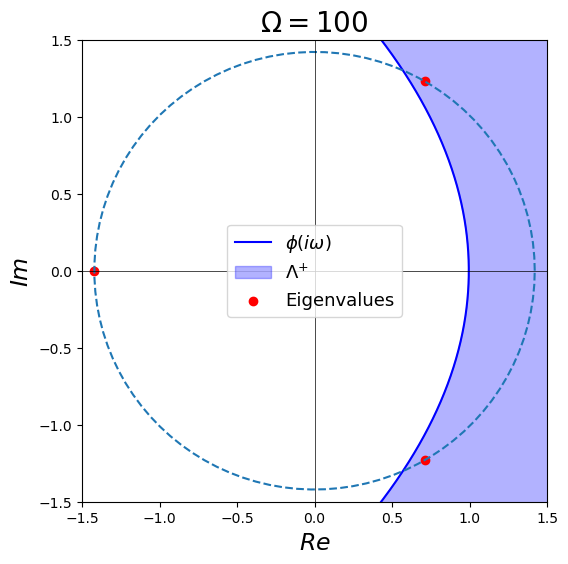}\hfill
\caption{\textbf{Graphical criteria} }
  \end{subfigure}\par\medskip
  \begin{subfigure}{\linewidth}
\includegraphics[width=.33\textwidth]{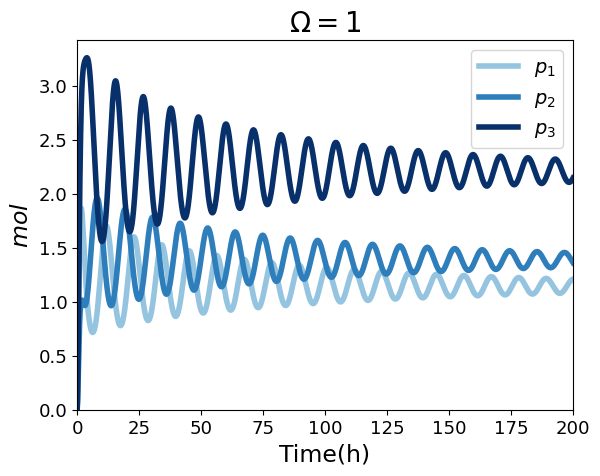}\hfill
\includegraphics[width=.33\textwidth]{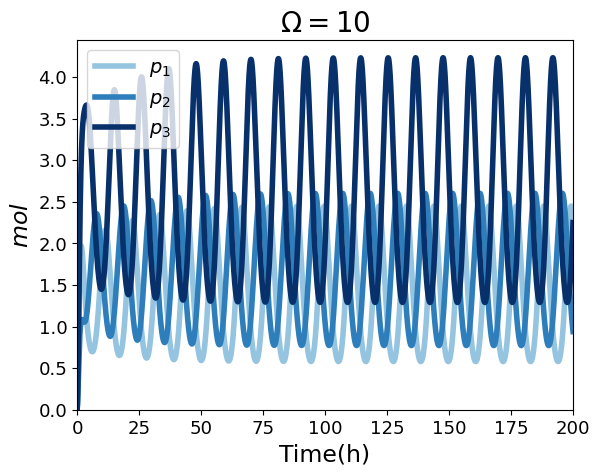}\hfill
\includegraphics[width=.33\textwidth]{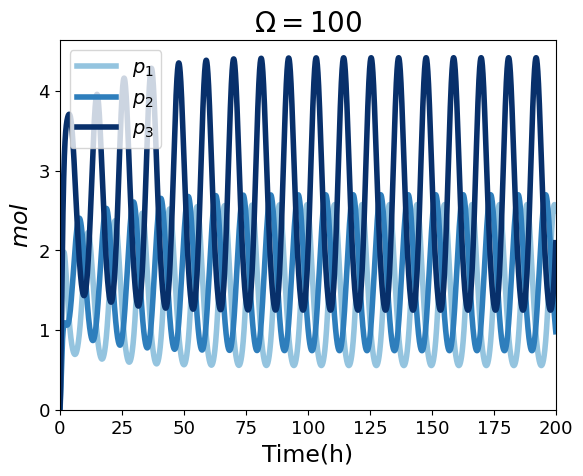}\hfill
\caption{\textbf{Dynamic of mean concentration of proteins} }
  \end{subfigure}
\caption{\textbf{Criteria of oscillation in the repressilator.} In this figure, we show that whether $\Omega$ increases, the system oscillates, which is based on graphical criteria. In panel (a), the results of the \MH{proposition 2} show that when $\Omega$ increases, the eigenvalues reach the region $\Lambda^{+}$, and then the system oscillates.  In panel (b), we observe the behavior of the oscillation dynamics as $\Omega$ increases (we used the initial conditions of Table  \ref{tabla3}). } 
\label{fig.3}
\end{figure*}

To analyze this system, we used $n=3$,  $\bm{\gamma}^2=(8,4,8)$, $\delta=1$,  and different values for $\Omega$ to explore the impact of size on the dynamics of the system; the values of the stationary mean concentrations of mRNA are listed in Table \ref{tabla1}, note that when $\Omega$ (the size of the system) decreases then decreases the stationary values. The values of the other variables were obtained using Equations (\ref{10}) and (\ref{11}) by substituting the value of the mRNA.

We first find the transfer function using Equation \eqref{20} and calculate the eigenvalues of $\mathbf{K}_1$ for different values of the system size $\Omega$; thus, we evaluate Proposition 2, as shown in Figure \ref{fig.3}.  These results indicate that when the size of the system $\Omega$ decreases, the system dynamics are governed by fluctuations, and oscillations do not appear. Thus, as fluctuations increase, the oscillations in the system are attenuated.

\subsection{\MH{Penta-silator}}

\begin{figure}[h!t]
    \centering
    \includegraphics[width=0.5\linewidth]{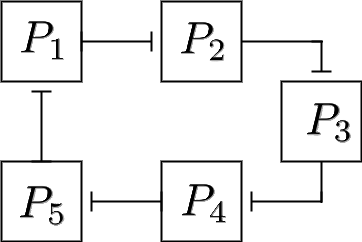}
    \caption{\textbf{\MH{Penta-silator}.} This is a cycle gene regulatory network with five modules in which each module represses the synthesis of mRNA of the next.  }
    \label{fig.4}
\end{figure}

\begin{figure*} [h!t]
  \begin{subfigure}{\linewidth}
\includegraphics[width=.33\textwidth]{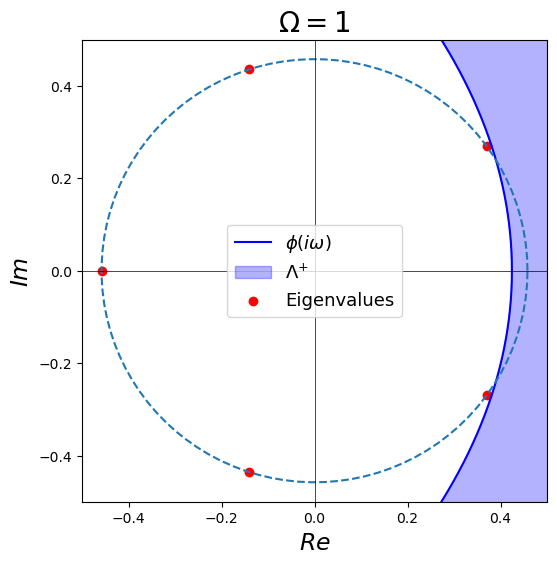}\hfill
\includegraphics[width=.33\textwidth]{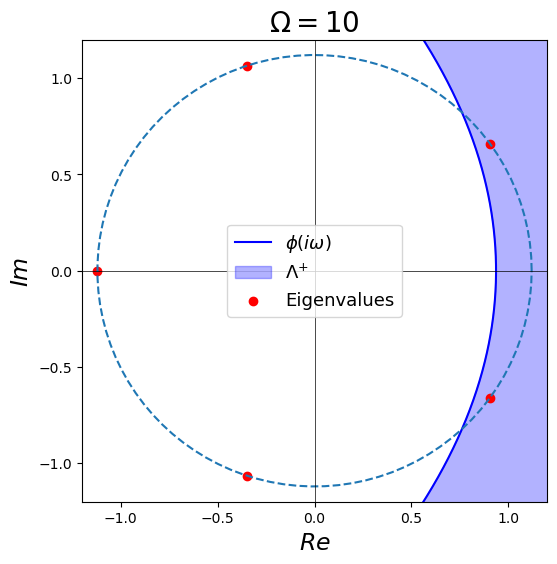}\hfill
\includegraphics[width=.33\textwidth]{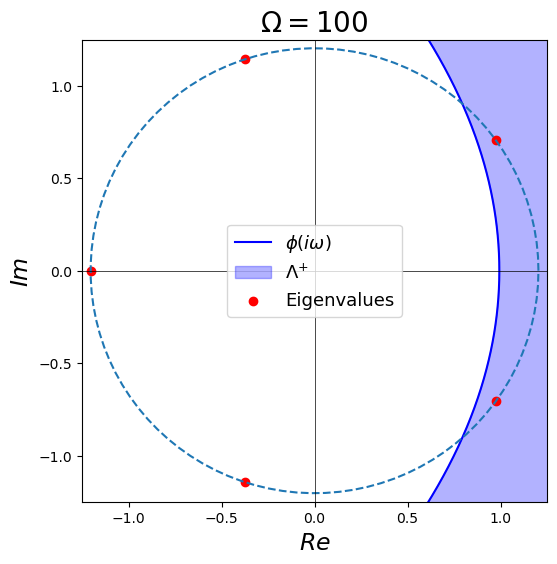}\hfill
\caption{\textbf{Graphical Criteria} }
  \end{subfigure}\par\medskip
  \begin{subfigure}{\linewidth}
\includegraphics[width=.33\textwidth]{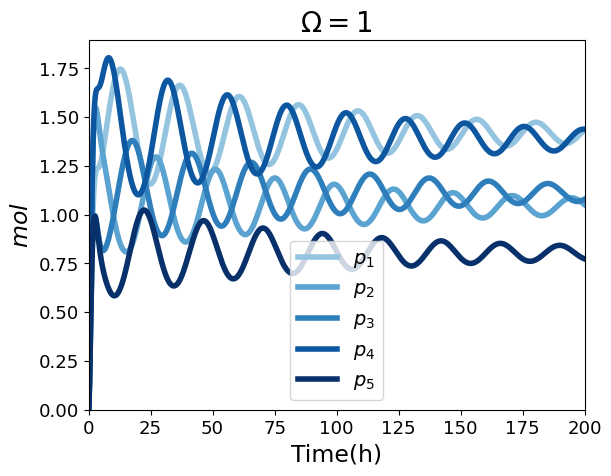}\hfill
\includegraphics[width=.33\textwidth]{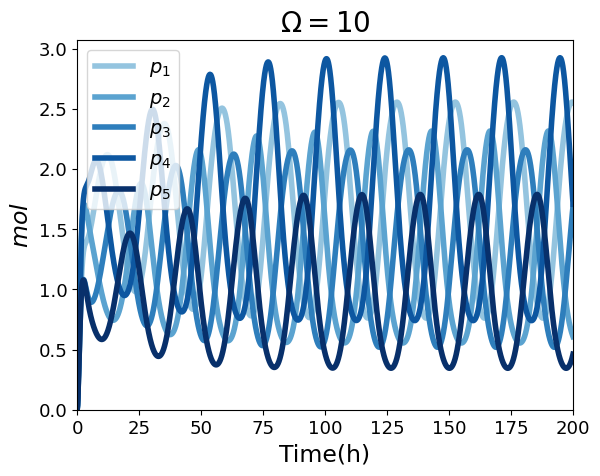}\hfill
\includegraphics[width=.33\textwidth]{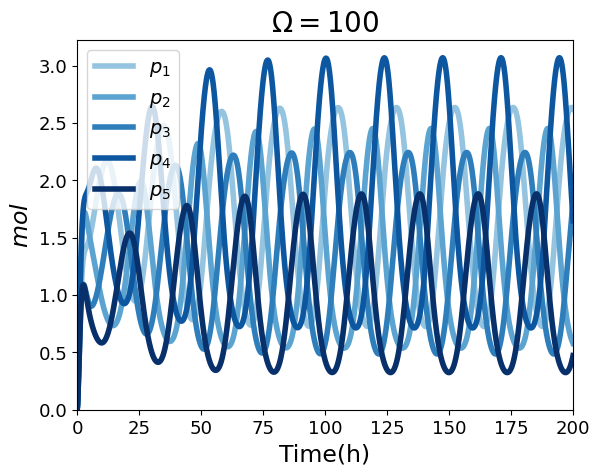}\hfill
\caption{\textbf{Dynamic of mean concentration of proteins} }
  \end{subfigure}
\caption{\textbf{Criteria of oscillation in the \MH{penta-silator}.} In this figure, we show whether $\Omega$ increases, the system oscillates; this criterion is based on graphical criteria. The mean concentrations are shown in panel (a). It can be seen that if $\Omega$ increases, the eigenvalues reach the region $\Lambda^{+}$, and then the system oscillates.  In panel (b), we observe the behavior of oscillation dynamics as $\Omega$ increases (we used the initial conditions of Table  \ref{tabla3}).} 
\label{fig.5}
\end{figure*}

The next system analyzed a gene regulatory network with a cycle, comprising five transcription-translation modules, as shown in Figure \ref{fig.4}. \MH{This system was developed experimentally by \cite{Meyer}.} We refer to this system as a penta-silator. We assume that the dynamics of the system are described by Equations \eqref{8}. To analyze this system, we used $\bm{\gamma}^2=(3,4,3,4,3)$, $\delta=0.7$,  and we use different values for $\Omega$ to explore the impact of size on the dynamics of the system; the values of the stationary mean concentrations of mRNA are listed in Table \ref{tabla2} and we observe that when the size of the system decreases the value of the stationary mean decreases. The values of the other variables were obtained using Equations (\ref{10}) and (\ref{11}) by substituting the value of the mRNA.

\begin{table}[htbp]
\centering
\begin{tabular}{cccc}
\hline
\textbf{Variable} & \textbf{Values} & $\Omega$ \\
\hline
$\mathbf{m}_{ss}$ & $(1.416, 1.042,1.111, 1.385, 0.804)$  & 1 \\
$\mathbf{m}_{ss}$ & $(1.568, 1.126,1.285, 1.467, 0.926)$  & 10 \\
$\mathbf{m}_{ss}$ & $(1.585, 1.136,1.306, 1.475, 0.942)$  & 100 \\
\hline
\end{tabular}
\caption{\textbf{Stationary mean concentrations.} The stationary value of the mean concentration of the mRNA obtained with Equation \eqref{12} using different values of $\Omega$.  }
\label{tabla2}
\end{table}

We follow a similar procedure to the previous example to evaluate Proposition 2, as shown in Figure \ref{fig.5}.  These results indicate that when the size of the system $\Omega$ decreases, the system dynamics are governed by fluctuations, and oscillations do not appear. 

From the examples we present, we can include the system size $\Omega$ as an additional parameter that determines when oscillations appear, particularly relevant for small systems where fluctuations are relevant. \\

\textbf{Remark I}: \MH{When $\Omega = 1$, the system exhibits damped oscillations whose amplitude decreases over time until the stationary state is reached. As the number of modules increases, the rate of decay becomes slower, which may give the impression of sustained oscillations; however, the system still undergoes damped oscillatory behavior.} 

\textbf{Remark II}: \MH{The deterministic limit is recovered as $\Omega \to \infty$. For $\Omega = 100$, the system's dynamics closely resemble the deterministic description, as fluctuations are very small; therefore, the results are comparable to those of the deterministic case.}

\begin{table*}
\centering
\begin{tabular}{ccc}
\hline
\hline
\textbf{Variable} & \textbf{Description} & \textbf{Value}\\
\hline
$m_j(0)$& Initial mean concentration of \MH{mRNA}. & $0$ $mol$ \\
${p}_j(0)$& Initial mean concentration of \MH{proteins}. & $0$ $mol$ \\
$M^2_{m_j,m_j}(0)$& Initial second central moment of concentration \MH{mRNA}. & 0 $mol^2$ \\
$M^2_{m_j,p_j}(0)$& Initial second central moment of concentration \MH{mRNA} and proteins. & 0 $mol^2$ \\
$M^2_{p_j,p_j}(0)$& Initial second central moment of concentration proteins. & 0 $mol^2$ \\
\hline
\end{tabular} 
\caption{\textbf{Initial conditions.} Initial conditions for the system. ($j\in (1,2,3,4,5)$) In this table, we show the initial conditions used for the gene regulatory network with negative feedback.}  \label{tabla3}
\end{table*}

\textbf{Remark III}: \MH{In the examples considered, the oscillations vanished for small system sizes. However, as shown in the figures, although the spectral radius of the eigenvalues decreases as $\Omega$ decreases, the boundary of the instability region $\Lambda^+$ also shifts to the left. This suggests the existence of a regime in which oscillatory behavior may reemerge owing to the contribution of fluctuations. In other words, fluctuations can induce oscillation. }

\section{Results and Conclusions} \label{section5}

In this study, we investigated the conditions under which oscillations occur in gene regulatory networks with a large number of components. The high dimensionality of these systems poses significant analytical challenges. To address this, we extended prior deterministic approaches by incorporating intrinsic fluctuations via the second-moment method, which reformulates stochastic dynamics as a set of ODEs. Within this formulation, tools developed for ODEs (traditionally used in deterministic frameworks) can be applied to the analysis of stochastic systems.

The analysis further revealed that while systems affected by intrinsic fluctuations are observable, they are not fully controllable. In particular, the second central moment of the mRNA-protein concentration pair cannot be directly controlled; however, this variable remains stable, indicating that system stability depends on a subset of controllable variables. We also establish graphical criteria for detecting oscillations in Proposition 2.

We validated the framework using two representative examples: a repressilator comprising three modules and a five-module network composed solely of repressors, both subject to intrinsic fluctuations. In the repressilator, we found that decreasing the system size amplified intrinsic fluctuations and weakened oscillatory behavior, as the mean protein concentrations, which are strongly dependent on system size, govern the presence of oscillations. The size of the system is another parameter that depends on the oscillations and is particularly relevant for systems in which the size is small and dominates fluctuations in the system's dynamics. A similar trend is observed for the five-module system.

The methodology developed in this study applies to any gene regulatory network with a cyclic structure, negative feedback, and an arbitrary number of modules, even in the presence of intrinsic fluctuations. Although our study focused on systems closed at the second central moment, the framework can be extended to include higher-order moments, such as the third central moment, which is relevant for systems with Hill-type functions. Future work could also explore the influence of extrinsic fluctuations and examine whether the conditions for oscillations and controllability are preserved under more complex scenarios.

\section*{Acknowledgments}
Manuel E. Hernández-García acknowledges the financial support of SECITHI through the program "Becas Nacionales 2023".

\section*{Data}

\MH{The code used to compute the region of oscillations is provided at the following link: https://github.com/Hill137/graphical-criteria.git }

\section*{Declarations}
The authors declare no conflicts of interest regarding the publication of this article. 

All data generated or analyzed in this study are included in this published article.

\appendix

\section{Initial Conditions}

Table \ref{tabla3} lists the initial conditions of the proposed models presented in this study. Because both models presented in this document are very similar, we used $j \in (1,2,3,4,5)$. In the case of the \MH{repressilator}, we only used it until index $j=3$.

\section{\MH{Hill Function}} \label{A.0}

Here, we derive the Hill function that is used in the principal text; this derivation is based on \cite{Exact, Decimal}. For this, we have the following reactions 
\begin{eqnarray}
    R+2L \stackbin[k_{-}]{k_{+}}{\rightleftarrows} RL_2, \nonumber \\
    0 \stackbin[k_1]{k_2}{\rightleftarrows} L, \label{c.0}
\end{eqnarray}
The first reaction is the binding of $2$ ligands $L$ to receptor $R$ in a reversible process to form complex $RL_2$. The stoichiometric coefficients and the stoichiometric matrix are, 
\begin{align}
 \alpha_{ij}&= \begin{pmatrix}
		2& 1 & 0 \\
            0& 0 & 1  \\
            0& 0 & 0  \\
            1& 0 & 0
	\end{pmatrix} ,    &
 \beta_{ij}&=  \begin{pmatrix}
		0& 0 & 1 \\
            2& 1 & 0  \\
            1& 0 & 0  \\
            0& 0 & 0 
	\end{pmatrix}, \nonumber \\
 \Gamma_{ij}&= \begin{pmatrix}
		-2 & 2 & 1 & -1 \\
		-1 & 1 & 0 & 0 \\
		1  & -1 & 0 & 0 
	\end{pmatrix}. \label{c.1}
\end{align}
Let $L, R, S$ be the number of molecules of $L, R, RL_2$ respectively. Thus, the propensity rates are
\begin{align}
    a_1&= k_{+} R \frac{L!}{(L-2)!} \frac{1}{\Omega^{3}}, & a_2=& k_{-} S\frac{1}{\Omega},     \label{c.2}
\end{align}
there is a conservative quantity $R+S=R_0$, the number of initial receptors, then the previous propensities are reduced to 
\begin{align}
    a_1&= k_{+} (R_0-S) \frac{L!}{(L-2)!} \frac{1}{\Omega^{3}}, & a_2=& k_{-} S\frac{1}{\Omega},     \label{c.3}
\end{align}
and 
\begin{align}
     \Gamma'_{ij}&= \begin{pmatrix}
		-2 & 2 & 1 & -1 \\
		1  & -1 & 0 & 0 
	\end{pmatrix},  
\end{align}
if we suppose that the first reactions in (\ref{c.0}) are in the stationary state, $S$ and $L$ are independent, then the central moments between $S$ and $L$ are zero, then we get,
\begin{align}
    \frac{\partial s}{\partial t}=0 = -k_{-} s + k_{+}(r_0 - s) \frac{1}{\Omega^{2}} \left \langle   \frac{L!}{(L-2)!}  \right \rangle,
\end{align}
where $K^2= \frac{k_{-}}{k_{+}}$,  $r_0=R_0/\Omega$, $s=\braket{S}/\Omega$ and $r=\braket{R}/\Omega$ are the mean concentrations of  $S$ and $R$, from this we get 
\begin{align}
    s= r_0 \frac{\frac{1}{\Omega^{2}}\left \langle   \frac{L!}{(L-2)!}  \right \rangle}{K^2+ \frac{1}{ \Omega^{2}}\left \langle   \frac{L!}{(L-2)!}  \right \rangle}.
\end{align}
If we define the Hill function as follows and substitute $r+s=r_0$ and the value of $s$, we have
\begin{align}
    H=\frac{s}{r+s}= \frac{s}{r_0} = \frac{\frac{1}{\Omega^{2}}\left \langle   \frac{L!}{(L-2)!}  \right \rangle}{K^2+ \frac{1}{ \Omega^{2}}\left \langle   \frac{L!}{(L-2)!}  \right \rangle}, \label{c.5}
\end{align}
using the second-moment framework and defined $l= \braket{L}/\Omega$ and $M^2_{l, l}= \braket{(L-\braket{L})^2}/\Omega^2$, we get
\begin{align}
        H= \frac{l^2 + M^2_{l,l} - \frac{l}{\Omega}  }{{K^2} + l^2 + M^2_{l,l} - \frac{l}{\Omega} }.
\end{align}
This expression is exact because we did not make any approximations in the derivation, and it is valid when the ligands bind and unbind to the receptors very fast. Deviations from the assumption of fast transcription factor binding and unbinding may compromise the validity of the Hill function approximation. In such cases, a more accurate description would require explicitly modeling the underlying biochemical reactions from which the Hill function was derived. However, this would result in a larger system of equations, thereby increasing the complexity and length of the analysis.

The derivation presented here is for an activator, but we can perform similar derivations for repressors or only use the relation $D=1-H$. For the recovery of the deterministic case, $\Omega \rightarrow \infty $ and $M^2_{l,l}=0$.

\section{Dimensionless} \label{A.A}

In this section, we describe how to reduce the number of parameters required to describe the system. In this case, we focus only on the ODEs for the mean concentrations because this procedure is very similar if we use all ODEs. 
\begin{subequations}
\begin{align}
    \frac{\partial m_i}{ \partial t}=& \gamma_{1,i} - \delta_1 m_i, \\
    \frac{\partial p_i}{ \partial t}=&  \gamma_{2,i} m_i - \delta_2 p_i ,   \\
    \gamma_{1,i}=& \gamma_{1,i}^* \left( \frac{K^2}{K^2 + p_{i-1}^2 + M^2_{p_{i-1},p_{i-1}}-\frac{p_{i-1}}{\Omega}}\right),
\end{align}
\end{subequations}
we observe that to describe the system we need $n$ parameters $\gamma_{1,i}$,  $n$ parameters $\gamma_{2,i}$, a parameter $\delta_1$, a parameter $\delta_2$, a parameter $K$ and a parameter $\Omega$, that is in total $2n+4$ parameters, to reduce this first we introduce new variables, and we substitute $m_i \rightarrow  m_{0,i} m_i$, $p_i \rightarrow p_{0,i} p_i$, $M^2_{p_{i},p_{i}} \rightarrow p_{0,i}^2 M^2_{p_{i},p_{i}}$, $\Omega \rightarrow \Omega_0 \Omega$ and $t \rightarrow \tau t$, and simplify we get
{\small
\begin{align}
    \frac{\partial m_i}{ \partial t}=& \frac{\tau}{m_{0,i}}  \gamma_{1,i} - \tau \delta_1 m_i, \nonumber \\
    \frac{\partial p_i}{ \partial t}=&  \frac{\tau m_{0,i}}{p_{0,i}} \gamma_{2,i} m_i - \tau \delta_2 p_i ,  \nonumber \\
    \gamma_{1,i}=& \gamma_{1,i}^* \left( \frac{K^2}{K^2 + p_{0,i-1}^2 p_{i-1}^2 + p_{0,i-1}^2 M^2_{p_{i-1},p_{i-1}}- \frac{p_{0,i-1}}{\Omega_0} \frac{p_{i-1}}{\Omega}}\right), \nonumber
\end{align}}
then we define $\tau= \frac{1}{\delta_2}$, $\delta=\frac{\delta_1}{\delta_2}$, $p_{0,i}= K$, $m_{0,i}= \frac{K \delta_2}{\gamma_{2,i}}$, $\Omega_0 = \frac{1}{K}$, $\gamma_i^2= \frac{\gamma_{1,i}^* \gamma_{2,i}}{K \delta_1 \delta_2}$, then we get 
\begin{subequations}
\begin{align}
    \frac{\partial m_i}{ \partial t}=& \delta( \gamma_i^* -  m_i) ,  \\
    \frac{\partial p_i}{ \partial t}=&  m_i -  p_i ,   \\
     \gamma_i^*=& \gamma_i^2 \left( \frac{1}{1+  p_{i-1}^2 + M^2_{p_{i-1},p_{i-1}}- \frac{p_{i-1}}{ \Omega}}\right),
\end{align}
\end{subequations}

from these equations we can see that only need $n$ parameters $\gamma^2_i$, a parameter $\delta$ and a parameter $\Omega$, that is $n+2$ parameters.  Although we use a repressive Hill function, it can be changed to an activating one.

\section{Kalman Decomposition} \label{B}

The system that we were analyzing is observable but not controllable, so to know if the system is stable, particularly if the variable is not controllable, we applied the Kalman decomposition \cite{Boley}. Because all modules have the same matrix $\mathbf{A}$, $\mathbf{B}$, and $\mathbf{C}$, we made the next analysis for an arbitrary module $j$. For this, we calculate the matrix of transformation, 

\begin{align}
    \mathbf{T}= \begin{pmatrix}
		1& 0 & 0 & 0 & 0  \\
            0 & 1 & 0& 0 & 0   \\
            1 & 0 & 0& 0 & 1   \\
            0 & 0 & 1& 0 & 0   \\
            0 & 0 & 0& 1 & 0   \\
	\end{pmatrix},
\end{align}
now, we calculate the matrix $\mathbf{A}_k= \mathbf{T}^{-1}\mathbf{AT}$ (where $\mathbf{T}^{-1}$ is the inverse of $\mathbf{T}$), then we get 
\begin{align}
    \mathbf{A}_k= \begin{pmatrix}
-\delta & 0 & 0 & 0 & 0 \\
1 & -1 & 0 & 0 & 0 \\
1 & 0 &  -(\delta + 1) & 0 &1 \\
1 & 1 & 2 & -2& 0 \\
0 & 0 & 0 & 0 & -2\delta \\
\end{pmatrix}
\end{align}
Note that this matrix has the form 
\begin{align}
    \mathbf{A}_k =\begin{pmatrix}
        A_{11} & A_{12} \\
        0  &   A_{22}
    \end{pmatrix}
\end{align}
with 
\begin{align}
    A_{11}=& \begin{pmatrix}
-\delta & 0 & 0 & 0  \\
1 & -1 & 0 & 0 \\
1 & 0 &  -(\delta + 1) & 0  \\
1 & 1 & 2 & -2 \\
\end{pmatrix}, \nonumber \\
A_{12}=& \begin{pmatrix}
    0 \\
    0 \\
    1 \\
    0
\end{pmatrix}, \nonumber \\
A_{22}=&-2 \delta
\end{align}
where the part $A_{11}$ is observable and controllable (OC), while the part $A_{22}$ is observable but not controllable (OnC), while $A_{12}$ is how the dynamic of the part OC impacts the dynamic of OnC. To ensure that the system is stable, the real part of the eigenvalues of $A_{22}$ is minus that zero, but this is satisfied because the eigenvalue is $-2\delta<0$, then the part OnC is stable. In this same way, the variables that are not controllable are $M^2_{m_j,p_j}$. 

Following the Kalman decomposition, we calculate the other matrix of the system, $\mathbf{B}_k=\mathbf{T}^{-1} \mathbf{B}$ and $\mathbf{C}_k= \mathbf{CT}$, then we get 
\begin{align}
    \mathbf{B}_k=& \begin{pmatrix}
\delta \\
0 \\
0 \\
 0 \\
 0
\end{pmatrix}, & \mathbf{C}_k=&  \begin{pmatrix}
0 & 1 & 0 & 0 & 0 \\
0 &  0 & 0 & 1 & 0\\
\end{pmatrix},
\end{align}
both matrix has the form
\begin{align}
    \mathbf{B}_k=& \begin{pmatrix}
B_{1} \\
0 
\end{pmatrix}, & \mathbf{C}_k=&  \begin{pmatrix}
C_{1} & 0
\end{pmatrix},
\end{align}
with 
\begin{align}
    B_{1}=& \begin{pmatrix}
\delta \\
0 \\
0 \\
 0 
\end{pmatrix}, & 
C_{1}= &  \begin{pmatrix}
0 & 1 & 0 & 0  \\
0 &  0 & 0 & 1 \\
\end{pmatrix},
\end{align}
where $C_1$ and $B_1$ correspond to the part observable and controllable. With this decomposition, we calculate again the transfer function like $\mathbf{g}(s)= C_{1} (s I-A_{11})^{-1} B_1$, we get the same transfer functions as in Equation \eqref{20}.

\section{\MH{Weinstein–Aronszajn Identity}} \label{D}

In this section, we present the Weinstein–Aronszajn identity (also known as Sylvester's determinant theorem) \cite{Pozrikidis}. For this, we use the following
\begin{subequations}
\begin{align}
    \mathbf{L}=& 
    \begin{pmatrix}
        I_{2n \times 2n} & \mathbf{G}(s) \\
        \mathbf{K} & I_{n \times n}
    \end{pmatrix} \nonumber \\
    =& \begin{pmatrix}
        I_{2n \times 2n} & \mathbf{G}(s) \\
        0 & I_{n \times n}
    \end{pmatrix} 
    \begin{pmatrix}
        I_{2n \times 2n}- \mathbf{G}(s)\mathbf{K} & 0 \\
        \mathbf{K} & I_{n \times n}
    \end{pmatrix} \label{23.b} \\
    =& \begin{pmatrix}
        I_{2n \times 2n} & 0 \\
        \mathbf{K} & I_{n \times n}
    \end{pmatrix}
    \begin{pmatrix}
        I_{2n \times 2n} & \mathbf{G}(s) \\
        0 & I_{n \times n} - \mathbf{K} \mathbf{G}(s),
    \end{pmatrix} \label{23.a} 
\end{align} 
\end{subequations}
now we calculi the determinant of $\mathbf{L}$, and because \eqref{23.a} and \eqref{23.b} are equal, we get
\begin{align}
    det(\mathbf{L})=& det(I_{2n \times 2n}- \mathbf{G}(s)\mathbf{K})  \nonumber \\
    =& det (I_{n \times n} - \mathbf{K} \mathbf{G}(s) ) \nonumber \\
    =& det( I_{n \times n} -  \mathbf{K}_1 g_1(s) -  \mathbf{K}_2 g_2(s)),
\end{align}
this is just the relation that we used in the principal text.


\nocite{*}
\bibliography{aipsamp}

@PREAMBLE{
 "\providecommand{\noopsort}[1]{}" 
 # "\providecommand{\singleletter}[1]{#1}%" 
}

@book{Alon,
  title={An introduction to systems biology: design principles of biological circuits},
  author={Alon, Uri},
  year={2019},
  publisher={Chapman and Hall/CRC}
}

@article{Gar,
  title={Handbook of stochastic methods for physics, chemistry, and the natural sciences Springer},
  author={Gardiner, CW},
  journal={Berlin (4th Ed)},
  year={2009}
}

@article{Gomez,
  title={Mass fluctuation kinetics: Capturing stochastic effects in systems of chemical reactions through coupled mean-variance computations},
  author={Gomez-Uribe, Carlos A and Verghese, George C},
  journal={The Journal of chemical physics},
  volume={126},
  number={2},
  year={2007},
  publisher={AIP Publishing}
}

@article{Gillespie,
  title={Exact stochastic simulation of coupled chemical reactions},
  author={Gillespie, Daniel T},
  journal={The journal of physical chemistry},
  volume={81},
  number={25},
  pages={2340--2361},
  year={1977},
  publisher={ACS Publications}
}

@article{Lakatos,
  title={Multivariate moment closure techniques for stochastic kinetic models},
  author={Lakatos, Eszter and Ale, Angelique and Kirk, Paul DW and Stumpf, Michael PH},
  journal={The Journal of chemical physics},
  volume={143},
  number={9},
  year={2015},
  publisher={AIP Publishing}
}

@article{Vitaterna,
  title={Overview of circadian rhythms},
  author={Vitaterna, Martha Hotz and Takahashi, Joseph S and Turek, Fred W},
  journal={Alcohol research \& health},
  volume={25},
  number={2},
  pages={85},
  year={2001}
}

@article{Mills,
  title={Human circadian rhythms.},
  author={Mills, JN},
  journal={Physiological reviews},
  volume={46},
  number={1},
  pages={128--171},
  year={1966}
}

@article{Schiavon,
  title={The art of modeling gene regulatory circuits},
  author={G{\'o}mez-Schiavon, Mariana and Montejano-Montelongo, Isabel and Orozco-Ruiz, F Sophia and Sotomayor-Vivas, Cristina},
  journal={NPJ Systems Biology and Applications},
  volume={10},
  number={1},
  pages={60},
  year={2024},
  publisher={Nature Publishing Group UK London}
}

@article{Xiao,
  title={Genetic oscillation deduced from Hopf bifurcation in a genetic regulatory network with delays},
  author={Xiao, Min and Cao, Jinde},
  journal={Mathematical biosciences},
  volume={215},
  number={1},
  pages={55--63},
  year={2008},
  publisher={Elsevier}
}

@inproceedings{Takada,
  title={Existence conditions for oscillations in cyclic gene regulatory networks with time delay},
  author={Takada, Masaaki and Hori, Yutaka and Hara, Shinji},
  booktitle={2010 IEEE International Conference on Control Applications},
  pages={830--835},
  year={2010},
  organization={IEEE}
}

@article{Walk,
  title={Information transmission in genetic regulatory networks: a review},
  author={Tka{\v{c}}ik, Ga{\v{s}}per and Walczak, Aleksandra M},
  journal={Journal of Physics: Condensed Matter},
  volume={23},
  number={15},
  pages={153102},
  year={2011},
  publisher={IOP Publishing}
}

@article{Lev,
  title={Generation of oscillations by the p53-Mdm2 feedback loop: a theoretical and experimental study},
  author={Lev Bar-Or, Ruth and Maya, Ruth and Segel, Lee A and Alon, Uri and Levine, Arnold J and Oren, Moshe},
  journal={Proceedings of the National Academy of Sciences},
  volume={97},
  number={21},
  pages={11250--11255},
  year={2000},
  publisher={The National Academy of Sciences}
}

@article{Extri,
  title={Extrinsic fluctuations in the p53 cycle},
  author={Hern{\'a}ndez-Garc{\'\i}a, Manuel Eduardo and G{\'o}mez-Schiavon, Mariana and Vel{\'a}zquez-Castro, Jorge},
  journal={The Journal of Chemical Physics},
  volume={161},
  number={18},
  year={2024},
  publisher={AIP Publishing}
}

@article{Smeenk,
  title={Characterization of genome-wide p53-binding sites upon stress response},
  author={Smeenk, Leonie and van Heeringen, Simon J and Koeppel, Max and van Driel, Marc A and Bartels, Stefanie JJ and Akkers, Robert C and Denissov, Sergei and Stunnenberg, Hendrik G and Lohrum, Marion},
  journal={Nucleic acids research},
  volume={36},
  number={11},
  pages={3639--3654},
  year={2008},
  publisher={Oxford University Press}
}

@article{Baroni,
  title={A global suppressor motif for p53 cancer mutants},
  author={Baroni, Timothy E and Wang, Ting and Qian, Hua and Dearth, Lawrence R and Truong, Lan N and Zeng, Jue and Denes, Alec E and Chen, Stephanie W and Brachmann, Rainer K},
  journal={Proceedings of the National Academy of Sciences},
  volume={101},
  number={14},
  pages={4930--4935},
  year={2004},
  publisher={National Academy of Sciences}
}

@article{Hirata,
  title={Oscillatory expression of the bHLH factor Hes1 regulated by a negative feedback loop},
  author={Hirata, Hiromi and Yoshiura, Shigeki and Ohtsuka, Toshiyuki and Bessho, Yasumasa and Harada, Takahiro and Yoshikawa, Kenichi and Kageyama, Ryoichiro},
  journal={Science},
  volume={298},
  number={5594},
  pages={840--843},
  year={2002},
  publisher={American Association for the Advancement of Science}
}

@article{Liu,
  title={Hes1: a key role in stemness, metastasis and multidrug resistance},
  author={Liu, Zi-Hao and Dai, Xiao-Meng and Du, Bin},
  journal={Cancer biology \& therapy},
  volume={16},
  number={3},
  pages={353--359},
  year={2015},
  publisher={Taylor \& Francis}
}

@book{VecchioM,
  title={Biomolecular feedback systems},
  author={Del Vecchio, Domitilla and Murray, Richard M},
  year={2015},
  publisher={Princeton University Press Princeton, NJ}
}

@article{Langevin,
  title={The chemical Langevin equation},
  author={Gillespie, Daniel T},
  journal={The Journal of Chemical Physics},
  volume={113},
  number={1},
  pages={297--306},
  year={2000},
  publisher={American Institute of Physics}
}

@misc{Scott,
  title={Applied stochastic processes in science and engineering},
  author={Scott, Matt},
  year={2013},
  publisher={Springer,}
}

@article{Exact,
author = {Hernández-García, Manuel Eduardo and Moreno-Barbosa, Eduardo and Velázquez-Castro, Jorge},
    title = {An exact moment-based approach for high-order chemical reaction–diffusion networks: From mass action to Hill functions},
    journal = {Chaos: An Interdisciplinary Journal of Nonlinear Science},
    volume = {36},
    number = {2},
    pages = {023122},
    year = {2026},
    month = {02},
    issn = {1054-1500},
    doi = {10.1063/5.0280665},
    url = {https://doi.org/10.1063/5.0280665}
}

@article{Manuel,
doi = {10.1088/2632-072X/ae3c4f},
url = {https://doi.org/10.1088/2632-072X/ae3c4f},
year = {2026},
month = {feb},
publisher = {IOP Publishing},
volume = {7},
number = {1},
pages = {015007},
author = {Hernández-García, Manuel Eduardo and Velázquez-Castro, Jorge},
title = {Fluctuation-induced corrections to the Hill function: implications for gene regulatory network dynamics},
journal = {Journal of Physics: Complexity}
}

@article{Decimal,
  title={Relationship between Decimal Hill Coefficient, Intermediate Processes, and Mesoscopic Fluctuations in Gene Expression},
  author={Hern{\'a}ndez-Garc{\'\i}a, Manuel Eduardo and Vel{\'a}zquez-Castro, Jorge},
  journal={ACS omega},
  volume={10},
  number={14},
  pages={13906--13914},
  year={2025},
  publisher={ACS Publications}
}

@article{Stability,
  title={Stability analysis under intrinsic fluctuations: a second-moment perspective of gene regulatory networks},
  author={Hern{\'a}ndez-Garc{\'\i}a, Manuel Eduardo and G{\'o}mez-Schiavon, Mariana and Vel{\'a}zquez-Castro, Jorge},
  journal={Physical Biology},
  volume={22},
  number={6},
  pages={066001},
  year={2025},
  publisher={IOP Publishing}
}

@article{Hori,
  title={Existence criteria of periodic oscillations in cyclic gene regulatory networks},
  author={Hori, Yutaka and Kim, Tae-Hyoung and Hara, Shinji},
  journal={Automatica},
  volume={47},
  number={6},
  pages={1203--1209},
  year={2011},
  publisher={Elsevier}
}

@inproceedings{Harat,
  title={Stability analysis of linear systems with generalized frequency variables and its applications to formation control},
  author={Harat, Shinji and Hayakawa, Tomohisa and Sugatat, Hikaru},
  booktitle={2007 46th IEEE Conference on Decision and Control},
  pages={1459--1466},
  year={2007},
  organization={IEEE}
}

@article{Elowitz,
  title={A synthetic oscillatory network of transcriptional regulators},
  author={Elowitz, Michael B and Leibler, Stanislas},
  journal={Nature},
  volume={403},
  number={6767},
  pages={335--338},
  year={2000},
  publisher={Nature Publishing Group UK London}
}

@inproceedings{Hanna,
  title={Repressilators and promotilators: Loop dynamics in synthetic gene networks},
  author={El Samad, Hana and Del Vecchio, Domitilla and Khammash, Mustafa},
  booktitle={Proceedings of the 2005, American Control Conference, 2005.},
  pages={4405--4410},
  year={2005},
  organization={IEEE}
}

@book{Smith,
  title={Monotone dynamical systems: an introduction to the theory of competitive and cooperative systems},
  author={Smith, Hal L},
  number={41},
  year={1995},
  publisher={American Mathematical Soc.}
}

@article{Kulasiri,
  title={A review of systems biology perspective on genetic regulatory networks with examples},
  author={Kulasiri, Don and Nguyen, Lan K and Samarasinghe, Sandhya and Xie, Zhi},
  journal={Current Bioinformatics},
  volume={3},
  number={3},
  pages={197--225},
  year={2008},
  publisher={Bentham Science Publishers}
}

@article{Ribeiro,
  title={A general modeling strategy for gene regulatory networks with stochastic dynamics},
  author={Ribeiro, Andre and Zhu, Rui and Kauffman, Stuart A},
  journal={Journal of computational Biology},
  volume={13},
  number={9},
  pages={1630--1639},
  year={2006},
  publisher={Mary Ann Liebert, Inc. 2 Madison Avenue Larchmont, NY 10538 USA}
}

@article{Banks,
  title={Stability of cyclic gene models for systems involving repression},
  author={Banks, HT and Mahaffy, JM},
  journal={Journal of Theoretical Biology},
  volume={74},
  number={2},
  pages={323--334},
  year={1978},
  publisher={Elsevier}
}

@article{Mallet,
  title={The Poincar{\'e}-Bendixson theorem for monotone cyclic feedback systems},
  author={Mallet-Paret, John and Smith, Hal},
  journal={Journal of Dynamics and Differential Equations},
  volume={2},
  number={4},
  pages={367--421},
  year={1990},
  publisher={Springer New York}
}

@article{Boley,
  title={Computing the Kalman decomposition: An optimal method},
  author={Boley, Daniel},
  journal={IEEE transactions on automatic control},
  volume={29},
  number={1},
  pages={51--53},
  year={2003},
  publisher={IEEE}
}

@article{Hara2013,
  title={Stability analysis of systems with generalized frequency variables},
  author={Hara, Shinji and Tanaka, Hideaki and Iwasaki, Tetsuya},
  journal={IEEE Transactions on Automatic Control},
  volume={59},
  number={2},
  pages={313--326},
  year={2013},
  publisher={IEEE}
}

@article{Meyer,
  title={Rapid cell-free forward engineering of novel genetic ring oscillators},
  author={Niederholtmeyer, Henrike and Sun, Zachary Z and Hori, Yutaka and Yeung, Enoch and Verpoorte, Amanda and Murray, Richard M and Maerkl, Sebastian J},
  journal={elife},
  volume={4},
  pages={e09771},
  year={2015},
  publisher={eLife Sciences Publications, Ltd}
}

@book{Pozrikidis,
  title={An introduction to grids, graphs, and networks},
  author={Pozrikidis, Constantine},
  year={2014},
  publisher={Oxford University Press}
}

\end{document}